\documentclass[twocolumn,preprint]{aastex63}
\usepackage{amsmath}
\usepackage{color}
\usepackage{makecell}
\usepackage{array}

\hypersetup{linkcolor=red,citecolor=blue}

\begin{document}
\title{\textbf{A Flare-related Decimetric Type-IV Radio Burst Induced by the X2 Radiation of Electron Cyclotron Maser Emission}}
\correspondingauthor{Yao Chen}\email{yaochen@sdu.edu.cn \\ $^{\dagger}$These authors contributed equally to this work.}

\author[0000-0000-0000-0000]{Maoshui Lv$^{\dagger}$}
\affil{Institute of Space Sciences, Shandong University, Weihai, Shandong, 264209, China\\}

\author[0000-0001-5483-6047]{Ze Zhong$^{\dagger}$}
\affil{Center for Integrated Research on Space Science, Astronomy, and Physics, Institute of Frontier and Interdisciplinary Science, Shandong University, Qingdao 266237, China\\}
\affil{Institute of Space Sciences, Shandong University, Weihai, Shandong, 264209, China\\}

\author[0000-0003-1034-5857]{Xiangliang Kong$^{\dagger}$}
\affil{Institute of Space Sciences, Shandong University, Weihai, Shandong, 264209, China\\}
\affil{Center for Integrated Research on Space Science, Astronomy, and Physics, Institute of Frontier and Interdisciplinary Science, Shandong University, Qingdao 266237, China\\}

\author[0000-0001-8132-5357]{Hao Ning}
\affil{Center for Integrated Research on Space Science, Astronomy, and Physics, Institute of Frontier and Interdisciplinary Science, Shandong University, Qingdao 266237, China\\}
\affil{Institute of Space Sciences, Shandong University, Weihai, Shandong, 264209, China\\}

\author[0000-0002-1576-4033]{Feiyu Yu}
\affil{Institute of Space Sciences, Shandong University, Weihai, Shandong, 264209, China\\}

\author[0000-0003-4025-0525]{Bing Wang}
\affil{Institute of Space Sciences, Shandong University, Weihai, Shandong, 264209, China\\}

\author[0000-0003-2047-9664]{Baolin Tan}
\affil{CAS Key Laboratory of Solar Activity, National Astronomical Observatories, Chinese Academy of Sciences, Beijing, 100012, China\\}

\author[0000-0002-7357-211X]{Victor Melnikov}
\affil{Central Astronomical Observatory at Pulkovo of the Russian Academy of Sciences, St Petersburg, 196140, Russia\\}

\author[0000-0001-8644-8372]{Alexey Kuznetsov}
\affil{Institute of Solar-Terrestrial Physics, Irkutsk, 664033, Russia\\}

\author[0000-0001-5705-661X]{Hongqiang Song}
\affil{Institute of Space Sciences, Shandong University, Weihai, Shandong, 264209, China\\}

\author[0000-0002-2734-8969]{Ruisheng Zheng}
\affil{Institute of Space Sciences, Shandong University, Weihai, Shandong, 264209, China\\}

\author[0000-0001-6449-8838]{Yao Chen}
\affil{Center for Integrated Research on Space Science, Astronomy, and Physics, Institute of Frontier and Interdisciplinary Science, Shandong University, Qingdao 266237, China\\}
\affil{Institute of Space Sciences, Shandong University, Weihai, Shandong, 264209, China\\}

\begin{abstract}
The radiation mechanism of decimetric wideband and pulsating radio bursts from the Sun (in terms of decimetric type-IV (t-IVdm) burst) and other flaring stars is a long-standing problem. Early investigations were based on the leading-spot hypothesis for the sun and yielded contradictory results. Here, we analyzed the flare-associated t-IVdm burst on 20110924 with medium-strong levels of polarization and from sources near a sunspot. We found that the emission is intermittent and the maximum $T_B$ exceeds 10$^{11}$ K, with well-defined upper and lower frequency cutoffs. The radio sources are left-handed polarized, located above the sunspot with a negative polarity. The sources align well with the sites of the second harmonic of the local electron gyrofrequency. These findings provide essential evidence that the burst is induced by the electron cyclotron maser emission (ECME) in the harmonic X mode. We further modeled the transport of downward-streaming energetic electrons along a coronal loop and found most electrons get mirrored within the specific altitude range of 20--100 Mm. This explains why such bursts tend to have well-defined spectral ranges. We also found the ECME-radiating energetic electrons exhibit a shell-like VDF instead of the generally-presumed loss-cone distribution. The study greatly expands the application of ECME in solar radio astronomy and provides solar samples for similar bursts from other flaring stars.
	
\end{abstract}

\keywords{Solar radio emission (1522) --- Solar flares (1496) --- Solar activity (1475) --- Solar magnetic fields (1503)}

\section{Introduction} \label{sec:intro}
Decimetric broadband radio bursts (with frequencies of hundreds of MHz to a few GHz) have been observed from the Sun \citep{1961Young,1970DeGroot,1972Gotwols} and other flaring stars \citep{1989Gudel,1990Lang,1990Bastian}. In the solar case, such bursts have been classified as the decimetric type-IV bursts (t-IVdm). The bursts tend to have well-defined low and high-frequency cutoffs and are closely associated with flares. Their spectra often contain substantial fine structures, such as pulsations, spikes, zebras, fibers, and various absorption structures. This makes them distinct from their metric and centimetric counterparts.

Due to the close proximity of the Sun to the Earth, solar radio astronomy has the advantage of observations with high spectral and spatial resolutions. This favors the determination of the underlying radiation mechanism. Even so, the exact origin of solar t-IVdm bursts and their substantial fine structures remain largely unresolved \citep{1980Bernold,1980Benz,1982Slottje,1985Dulk,1988Guedel,1994Isliker,2005Fleishman,2016Vasanth,2019Morosan,2024White}.

To determine the primary radiation mechanism, it is essential to identify the nature of the emission mode in terms of two freely escaping magnetoionic modes: extraordinary (X) or ordinary (O). This requires knowledge of the sense of polarization and orientation of the source magnetic field. Due to the lack of imaging data at decimetric wavelengths, to determine the source field orientation earlier studies \citep{1986Aschwanden,1987Zlobec} were based on the leading sunspot hypothesis that assumes the bursts originate from the leading spots of the associated flares. Contradictory conclusions were drawn by these studies. As a result, distinct mechanisms have been proposed: plasma emission \citep[PE;][]{1974Kuijpers,1980Kuijpers,1975Zaitsev,1976aBenz, 1985Dulk}, electron cyclotron maser emission \citep[ECME;][]{1976bBenz,1988aAschwanden,1988bAschwanden}, gyrosynchrotron radiation (e.g., \citealt{1973Dulk,1985Dulk,1994Isliker,2019Morosan}), or resonant transition radiation \citep{2005Nita, 2005Fleishman} 

Over the past eight decades of solar radio astronomy, only solar spikes that are highly transient (ms), highly-polarized, and extremly bright (with the brightness temperature ($T_B$) reaching above 10$^{13}$ K) are widely attributed to ECME, while several other types of coherent solar bursts such as type-I, II, and III have been attributed to PE (e.g., \citealt{1985Dulk}; \citealt{1991Guedel}; \citealt{1998Fleishman}; \citealt{ 2024White}). Note that some authors also suggested PE to be the origin of certain spike-like emission (e.g., \citealt{2001Abalde}; \citealt{2001Chernov}). Recent studies have reported a rare ECME-like event with week-long decimetric pulsations, which is not directly associated with any specific flare \citep{2024Yu,2024Sharma}. Here, we analyzed flare-induced t-IVdm bursts using multiwavelength data to identify the underlying radiation mechanism. In Section \ref{s:obs}, we introduce how we selected the events, then we present the modeling methods of electron transport in Section \ref{s:mm}. The results are presented in Section \ref{s:results}, and the conclusions and discussion are provided in Section \ref{sec:discussion}.

\section{Observations and Data Analysis}\label{s:obs}
\subsection{Data and Instruments}\label{ss:Data}
The t-IVdm bursts were identified using the data from the 7m radio telescope of the Bleien observatory, which covers 175 to 870 MHz with a temporal resolution of 0.25 seconds. The Nan\c{c}ay Radioheliograph \citep[NRH;][]{1997Kerdraon} offers imaging data with polarization measurements at 10 frequencies spanning from 150 to 445 MHz. The spatial resolution depends on the frequency and time of observation, being $\sim$2$^{\prime}$ at 445 MHz and $\sim$6$^{\prime}$ at 150 MHz in the summer, and up to three times larger along the north-south direction in the winter.

The Solar Dynamics Observatory \citep[SDO;][]{2012Pesnell} provides EUV data with the Atmospheric Imaging Assembly \citep[AIA;][]{2012Lemen} and magnetograms with the Helioseismic and Magnetic Imager \citep[HMI;][]{2012Scherrer}. The EUV data have a pixel size of 0.6$^{\prime\prime}$ and a cadence of 12 s.

\subsection{Event Selection}\label{ss:Selection}
We collected t-IVdm bursts based on the data from the Bleien observatory in the frequency range of 175--870 MHz. To tell the location of the radio sources and the magnetic field orientation therein, we retained events observed during 2010--2014 when both radio and EUV imaging data were available and the radio data recorded by NRH were calibrated in both polarization and $T_B$.

According to the general definition of t-IVdm bursts, we selected events with the maximum frequency exceeding 300 MHz, the bandwidth wider than 100 MHz, and the duration longer than 5 minutes. This yielded 60 events, from which we selected 37 flare-associated events (Table~\ref{tab1}) according to the following criteria: the burst should be within 30 minutes around the peak of a flare, which originates from the active region (AR) associated with the NRH sources.

To identify the radiation mode, we only retained events with a high to moderate level of polarization ($>$30\%) whose sources are located near a sunspot, with both latitude and longitude within 60$^\circ$, to ensure an unambiguous determination of the source-field orientation. The field orientation is further verified by the nonlinear force-free field (NLFFF) extrapolation. After applying these criteria, we identified 2 events that occurred on 20110924 and 20120614, respectively. For the 20120614 event, the burst passes through the complex loop system of the AR, so we cannot determine the emission mode unambiguously due to the possible reversal of polarization (see Section \ref{sec:discussion}). Thus we only kept the 20110924 event here.

\section{Methods of Modeling Electron Transport Along the Loop}\label{s:mm}
\subsection{Magnetic Field Processing and Extrapolation}\label{ss:Extrapolation}
We analyzed the HMI vector magnetic field data (``hmi.B\_720s''), which cover the full solar disk with a temporal cadence of 720 s and a spatial resolution of 0.505$^{\prime\prime}$ \citep{2012Schou}. The coronal magnetic field was reconstructed using the magneto-frictional method \citep{2016Guo}. To do this, we first corrected the projection effect \citep{1990Gary} since the AR is away from the disk center, and then removed the residual force on the magnetograms using a preprocessing method \citep{2006Wiegelmann} with optimized parameters, ensuring that the extrapolation boundary remained consistent with observations. We applied two metrics \citep{2000Wheatland} to check whether the magnetic field reaches the force-free state or not. The obtained force-freeness metrics for both events are lower than 0.25, and the divergence-freeness metrics are close to 8 $\times$ 10$^{-5}$ , similar to previous studies \citep{2019Zhong}.

\subsection{Modeling of Electron Transport }\label{ss:Transport}
Following our previous studies \citep{2022Kong,2025Kong}, we modeled the transport of energetic electrons in the coronal loop by solving the focused transport equation numerically \citep{1969Roelof,1971Skilling,2018Effenberger}. Here, we only considered the magnetic mirroring effect and neglected the pitch-angle scattering by turbulence or Coulomb collision. The reduced particle transport equation is written as:
\begin{equation}
    \frac{\partial f}{\partial t}
    = -v \mu \hat{\mathbf{b}} \cdot \nabla f
    - \frac{v (1 - \mu^2)}{2 L_B} \frac{\partial f}{\partial \mu}
\end{equation}
where $f$ is the particle distribution function, $v$ is the particle speed, $\mu$ is the pitch-angle cosine, and $t$ is the time. The terms on the right-hand side describe the energetic electrons streaming along the direction of the magnetic field $\hat{\mathbf{b}}$ and the magnetic mirroring effect with the focusing length $L_B = \left( \hat{\mathbf{b}} \cdot \nabla \ln B \right)^{-1}$.

Since the above equation is basically a Fokker-Planck equation, it is mathematically equivalent to a set of stochastic differential equations \citep[SDEs;][]{1999Zhang,2017Strauss}. Here, we employed the time-forward SDEs to trace the particle's position and pitch-angle \citep{2022Kong,2025Kong}.

In the simulations, we assumed that the accelerated electrons have a Gaussian distribution in pitch-angle, $f(\mu) \sim \exp\left( -\frac{(\mu - 1)^2}{0.01} \right)$ with a power-law energy spectrum of $f(E) \sim E^{-3} $ with $E$ ranging from 10 to 100 keV. The electrons are impulsively released in the upper section of the selected loop, given by $x = [-484,-454]$ Mm, $y = [31,51]$ Mm, and $z = [150,170]$ Mm. We injected a total of 10 million pseudo-particles for the simulation. We obtained the 3D spatial distribution, pitch-angle and velocity-space distribution of energetic electrons at different times.

\section{Results}\label{s:results}
\subsection{Multiwavelength data analysis for the 20110924 event}\label{ss:20110924}
\begin{figure*}
\centering
\includegraphics[width=0.7\textwidth]{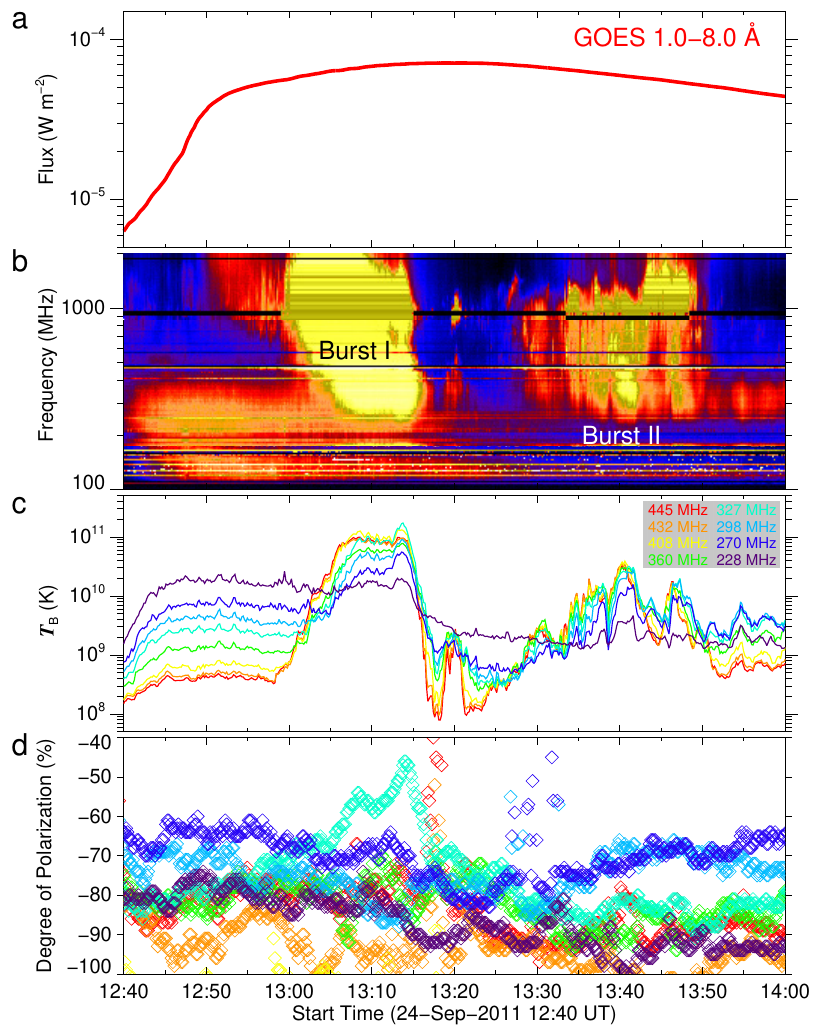}
\caption{
\textbf{Overview of the t-IVdm radio burst observed on 24 September 2011.} \textbf{(a)} GOES soft X-ray fluxes of the M7.1-class flare at 1.0--8.0 {\AA}. \textbf{(b)} The dynamic spectra observed from 12:40 to 14:00 UT, combining the data from the San Vito observatory (100--175 MHz), the Bleien observatory (175--870 MHz), and the Ondrejov observatory (870--2000 MHz). Bursts I and II are two components of the t-IVdm event. \textbf{(c)} The temporal profiles of the maximum $T_B$ and the degree of polarization \textbf{(d)} with a 10s cadence at 8 NRH frequencies from 228 to 445 MHz. The polarization level is obtained by taking average over the 85\% contours of the maximum $T_B$.
}
\label{fig1}
\end{figure*}

Fig.~\ref{fig1} presents the spectrum of the t-IVdm burst, along with the NRH data for $T_B$ and polarization. The event occurred from 12:50 to 13:50 UT and consists of two major sub-bursts (Bursts I and II), each lasting $\sim$20 minutes. The observations reveal three features: (1) Both burst components are highly intermittent, with well-defined upper and lower frequency cutoffs at $\sim$1--2 GHz and 200--300 MHz, respectively. (2) The maximum $T_B$ is $\sim$2$\times$10$^{11}$ K for Burst I and $\sim$4$\times$10$^{10}$ K for Burst II. (3) The polarization is left-handed with levels reaching $\sim$70--100\%, with an overall increasing trend with frequency. Based on these observations, we conclude that the emission is coherent since the incoherent gyrosynchrotron radiation of solar flares cannot yield such intermittency with high $T_B$ $>$10$^{11}$ K and strong polarization at such frequencies \citep{1985Dulk,2004Gary}.

\begin{figure*}
\centering
\includegraphics[width=1.0\textwidth]{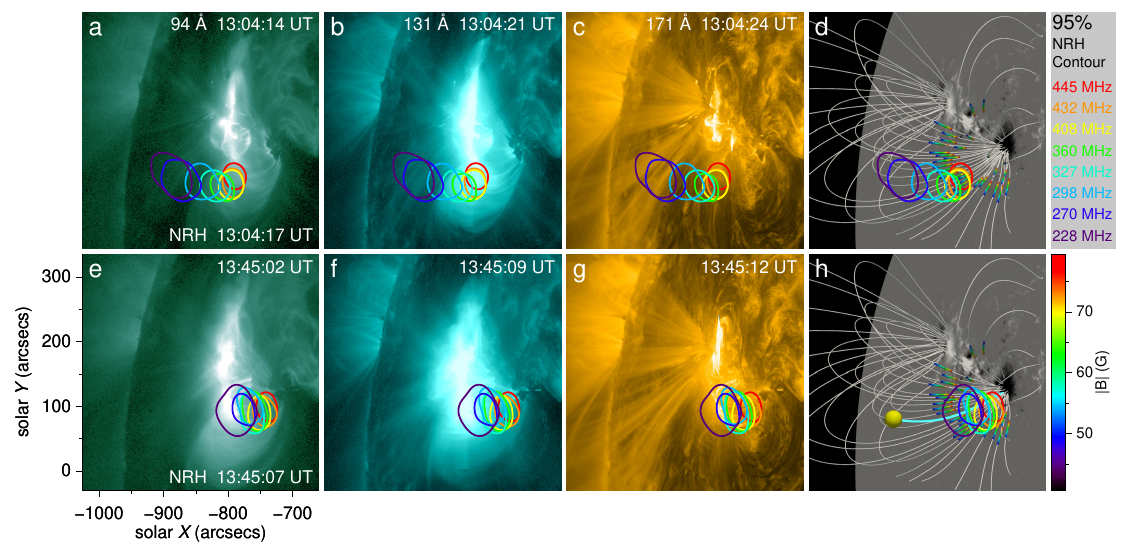}
\caption{
\textbf{The AIA and HMI data of the event.} \textbf{(a--c)} and \textbf{(e--g)} The AIA images at 94, 131, and 171 {\AA} are overlaid with NRH 95\% contours of the maximum $T_B$ observed at 8 frequencies (228--445 MHz) for representitive moments of bursts I and II. \textbf{(d)} and \textbf{(h)} The HMI magnetogram superposed by extrapolated NLFFF field lines. The colored sections of field lines correspond to the field strengths increasing from $\sim$40.7 to $\sim$79.5 G and the corresponding 2 $\Omega_{ce}$ increasing from 228 to 445 MHz. Colored circles display NRH 95\% contours at 13:04:17 \textbf{(d)} and 13:45:07 UT \textbf{(h)}. The yellow sphere crossing the cyan field line marks the location where energetic electrons are injected for the particle-transport simulation. An animation of this figure is available. It presents the temporal evolution of the radio sources and EUV structures from 12:40 UT to 14:00 UT.
}
\label{fig2}
\end{figure*}

Fig.~\ref{fig2} and the accompanying animation present the eruption recorded by AIA at 94, 131, and 171 {\AA}, superposed by the NRH source contours at 8 frequencies from 228 to 445 MHz. This event was associated with an M7.1-class flare originating from NOAA AR 11302. The flare started at 12:33 UT, peaked at 13:17 UT, and ended at 14:10 UT. The AR has a complex multipolar topology with three sunspots. The event was associated with a halo coronal mass ejection. Both 94 and 131 {\AA} data show the eruption of a hot structure and the rise of heated post-flare loops \citep{2023Fu}.

The NRH sources moved systematically during the burst, exhibiting at least two leaps since 12:50 UT. These leaps indicated the start of bursts I and II. Around 13:00 UT, the sources moved to the region above the flaring loops, and at $\sim$13:20 UT, they leaped toward the western leg of the loops, right above the rightmost sunspot with a negative polarity. During the whole process, the NRH sources aligned well with each other, and higher-frequency sources are closer to the disk.

Using the NLFFF extrapolation method, we deduced the coronal field lines of the AR (Figs.~\ref{fig2}d and \ref{fig2}h), and overlaid them onto the HMI magnetogram. The field lines are color-coded to represent the section where the field strength varies from $\sim$40.7 G (blue) to $\sim$79.5 G (red), the corresponding harmonic gyrofrequency (2 $\Omega_{ce}$) increases from 228 to 445 MHz.

We draw two major conclusions: (1) NRH sources of both bursts (I and II) lie along the field lines pointing toward the sunspot with strong left-handed polarization (Fig.~\ref{fig1}c), so both bursts are of X mode, and (2) the sources align well with the colored 2 $\Omega_{ce}$ section. Thus, the most likely radiation mechanism is the harmonic X mode (X2) via ECME since the alternative coherent plasma emission process would produce O mode for the fundamental branch or weak polarization for the harmonic branch \citep{1985Dulk,1978Melrose,2017Melrose,2020Ni,2022Chen,2022Zhang}.

\subsection{Possibility of Polarization Reverse for the 20110924 event}\label{ss:polarization}
According to the theory of mode coupling, the polarization sense of the radiation will be reversed when it propagates through a weak-coupling quasi-transverse (QT) layer, i.e., the radiation frequency is lower than the transition frequency of the QT layer \citep{1960Cohen, 1977Benz, 1992White}.

\begin{figure*}
	\centering
	\includegraphics[width=1.0\textwidth]{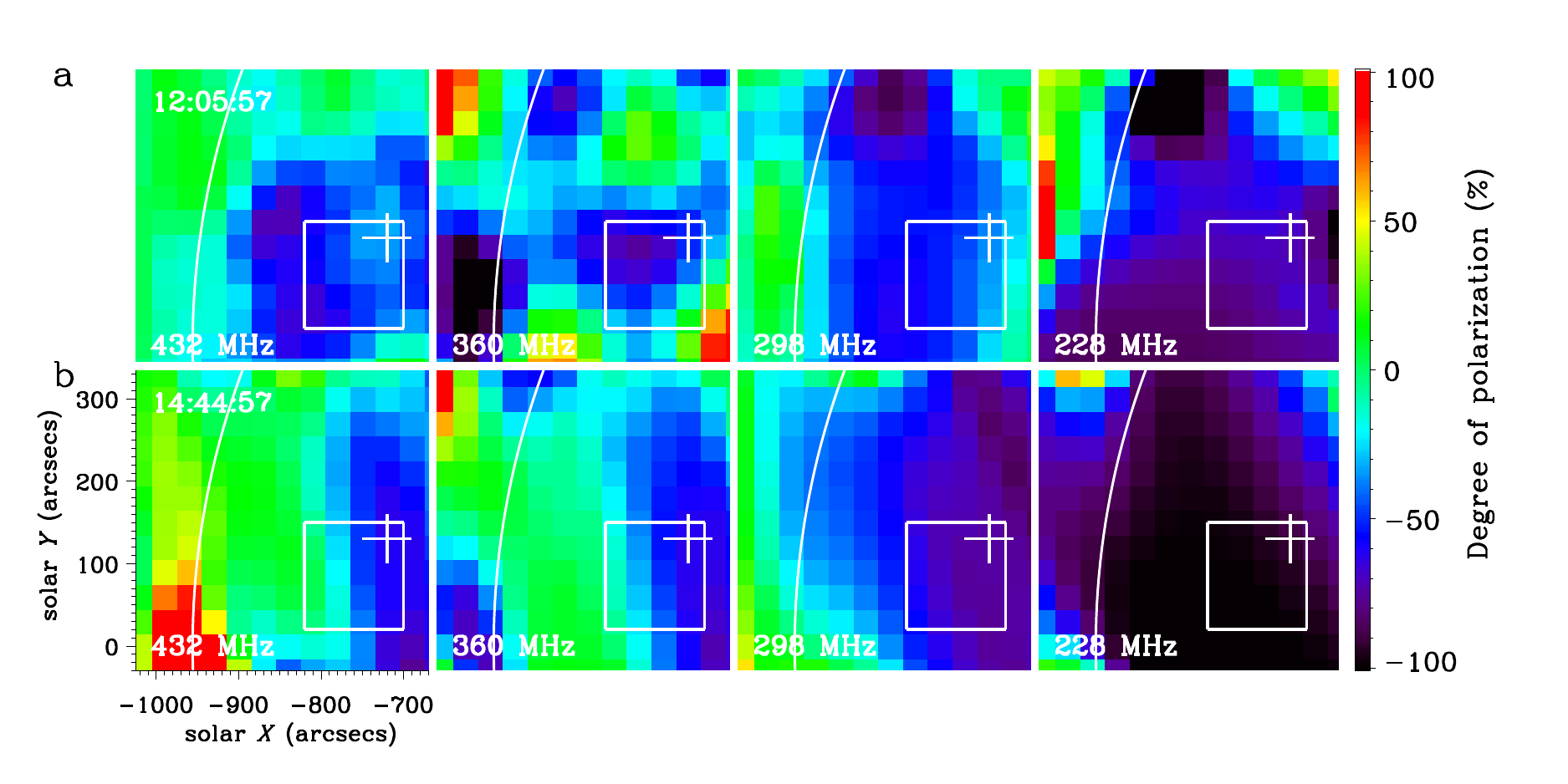}
	\caption{
		\textbf{The AR polarization at 4 NRH frequencies before (a) and after (b) the t-IVdm burst.} The box indicates where the radio sources were observed during the burst, and the plus indicates the sunspot location close to the radio sources.		
	}
	\label{fig3}
\end{figure*}

To determine whether the polarization of the burst is reversed by the propagation effect, we checked the AR polarization before and after the burst (Fig.~\ref{fig3}). The polarization in the burst source is generally left-handed, with the same sense to the burst. The non-eruptive AR radiates in radio via gyroresonance and bremsstrahlung of thermal electrons, mainly in the X mode (see, e.g., \citealt{1985Dulk}; \citealt{1997White}; \citealt{2004Gary}).

We further examined the height of the ejecta during the burst. According to the Large Angle and Spectrometric Coronagraph (LASCO C2; \citealt{1995Brueckner}), the coronal mass ejection as a whole has moved beyond 2 $R_s$ above the solar surface during the burst. This indicates the complex eruptive structure has moved beyond the weak-coupling regime that is roughly below 0.9 $R_s$ for the current event. This deduction agrees with AIA data of the eruption. So during the burst, the complex eruptive structure cannot result in the sense reversal of polarization. Combining the above analyses, we suggest that the burst radiation is also in X mode.

\subsection{Modeling electron transport for the 20110924 event}\label{ss:particle}
Magnetic field lines converge toward the sunspot as their strength increases, forming a magnetic mirror within which energetic electrons can be reflected. Around the mirroring point, the electron pitch angle ($\theta$) is $\sim$90$^\circ$ and the kinetic energy is mainly carried by the perpendicular motion. At locations with sufficient‌ mirrored electrons, the electron velocity distribution function (VDF) shall develop significant positive gradient (i.e., $\partial f/\partial v_{\perp} > 0$), which can excite ECME to radiate preferentially in X mode \citep{1958Twiss,1979Wu,2021Ning,2021Yousefzadeh}.

\begin{figure*}
\centering
\includegraphics[width=0.9\textwidth]{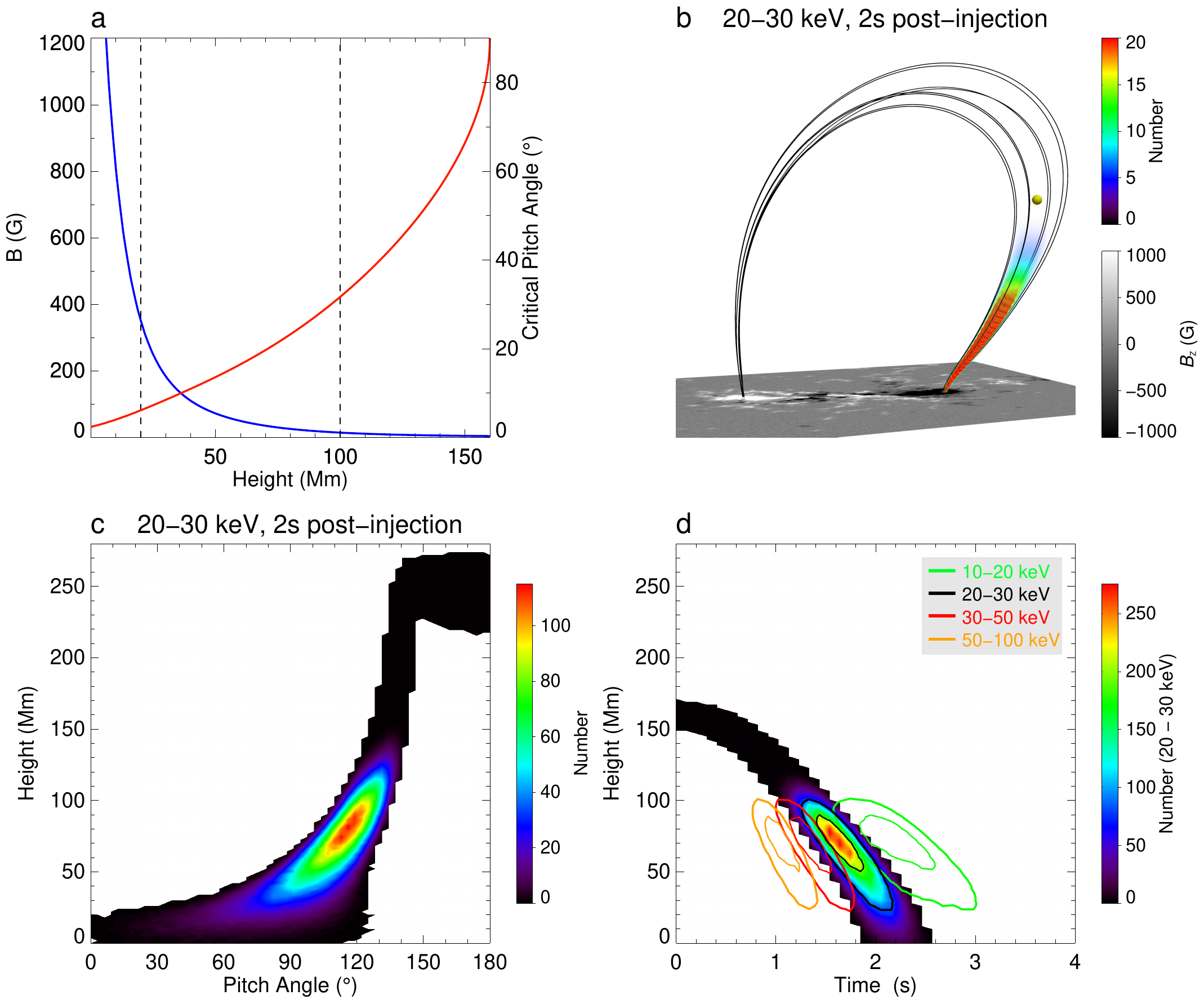}
\caption{
\textbf{Modeling results of the particle transport along the selected magnetic loop.} \textbf{(a)} Blue and red lines show the profiles of the magnetic field strength and critical pitch angles $\theta_c$ along the selected cyan field line plotted in Fig. 2\textbf{(h)}. The two vertical dashed lines delineate altitude range of 20 to 100 Mm. \textbf{(b)} The 3D spatial distribution at different altitudes (along the $z$-direction) of 20--30 keV electrons, 2s post-injection. The bottom map represents the HMI magnetogram. The yellow sphere marks the injection spot (see Fig. 2\textbf{(h)}). \textbf{(c)} The pitch angle distribution of 20--30 keV electrons versus the alitutude ($z$) above the solar surface. \textbf{(d)} The altitudes of mirroring points (where the pitch angle $\theta$ = 90$^\circ$) of 20--30 keV electrons as a function of time, overplotted with the 30\% and 70\% contours of the respective maxima, contours at different energy bands are also plotted. Approximately 89.7\% of electrons are reflected between 20 to 100 Mm for all energy ranges. An animation of panel \textbf{(b)} is available. It presents the dynamic evolution of the spatial distribution of energetic electrons along the loop.
}
\label{fig4}
\end{figure*}

To demonstrate whether the above scenario applies, we selected an extrapolated field line (the cyan line in Fig.~\ref{fig2}h) that clusters the NRH sources. The field strength $B$ increases rapidly with decreasing altitude (the blue line in Fig.~\ref{fig4}a), from $\sim$4.0 G at 160 Mm, to $\sim$14.5 G at 100 Mm, and reaching $\sim$351.8 G at 20 Mm. The critical pitch angles are given by $\theta_c = \arcsin \sqrt{B_0/B_m}$ (the red line in Fig.~\ref{fig4}a), where $B_0$ and $B_m$ are field strengths at the release and mirroring points, respectively. According to the basic theory of magnetic mirror, $\theta_c$ are 31.7$^\circ$ at 100 Mm and 6.1$^\circ$ at 20 Mm. This large change of $\theta_c$ implies a significant population of downward-propagating energetic electrons will get reflected within this range of altitude, as simulated below.

We assume the electrons are accelerated by the flare reconnection that takes place relatively high in the corona. The electrons then flow downward and upward. The Hard X-rays of solar flares and some drifting-toward-higher-frequency type-III bursts are evidence of these downward streaming energetic electrons (e.g., \citealt{1995Aschwanden}; \citealt{2014Reid}). To simulate the transport of these energetic electrons, we release a population of Gaussian-distributed beam-like energetic electrons in the upper loop section (see the yellow sphere in Fig.~\ref{fig2}h at an altitude of 160 Mm). The electrons, with the distribution of $f(\mu) \sim \exp\left( -\frac{(\mu - 1)^2}{0.01} \right)$  where $\mu = \cos \theta$, first stream toward the negative sunspot and then get mirrored at different altitudes (Fig.~\ref{fig4}b and the animation~2). As shown in Fig.~\ref{fig4}c, a large fraction of 20--30 keV electrons get mirrored at heights above $\sim$10--20 Mm within 2 seconds after release, among them $\sim$85\% (97\%) have pitch angles exceeding 90$^\circ$ (60$^\circ$). Eventually, $\sim$89.7\% of all injected electrons are mirrored at heights between 20 and 100 Mm (Fig.~\ref{fig4}d). This result also holds for 10--20 keV, 30--50 keV, and 50--100 keV electrons since the mirroring point only depends on the initial pitch angles of electrons due to the neglect of wave/turbulence-particle interaction. In the real corona, the mirroring point depends on such interaction and other factors, such as variation of the magnetic field, Coulomb collisions, etc.

\begin{figure*}
\centering
\includegraphics[width=0.9\textwidth]{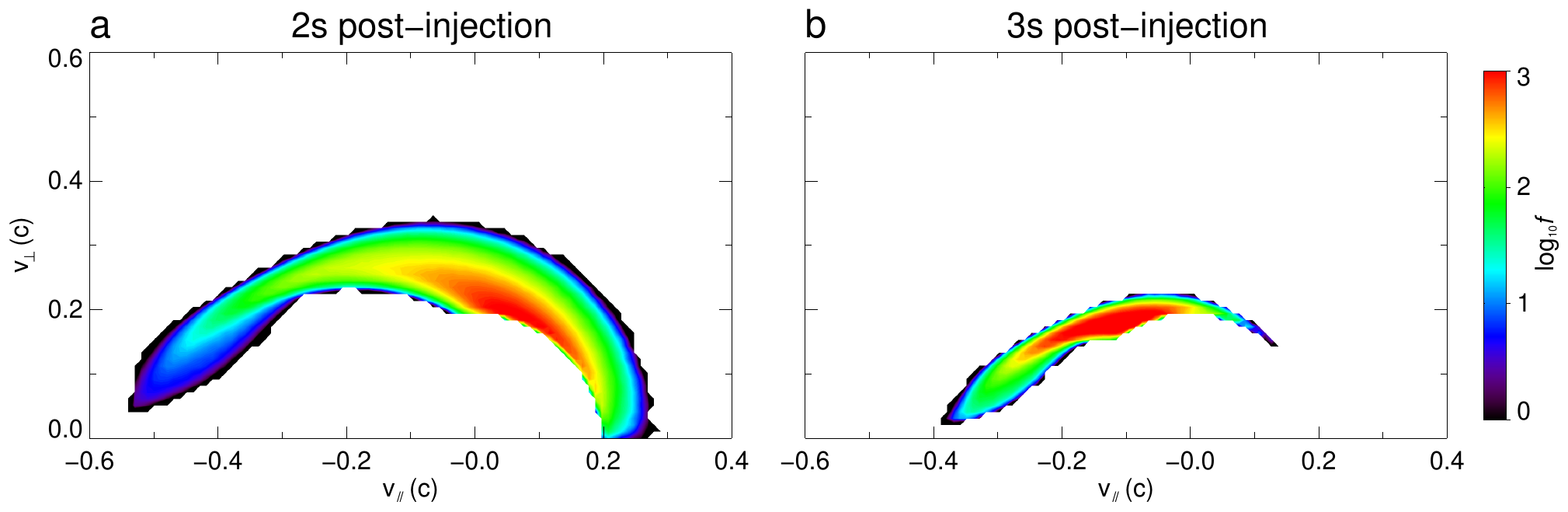}
\caption{
\textbf{Velocity distribution of energetic electrons at altitudes between 20 and 100 Mm at 2s (a) and 3s (b) post-injection, respectively.}
}
\label{fig5}
\end{figure*}

As expected, the electrons within 20--100 Mm exhibit a shell-like VDF (Fig.~\ref{fig5}), with significant positive gradient in the perpendicular direction. This type of VDF can drive the ECME to release both the fundamental O or X mode (O1 or X1) and X2. According to the latest particle-in-cell simulation \citep{2021Ning, 2021Yousefzadeh, 2022Yousefzadeh} with similar numerical setup, energetic electrons with such shell-like VDFs excite the X2 mode much more efficiently than the fundamental modes that will experience strong damping at the corresponding second harmonic absorption layer, while the absorption of X2 mode is much weaker \citep{1982Melrose,1989Robinson,1989McKean,2016Melrose}. This agrees with the above suggestion that the t-IVdm radiation is the X2 mode excited via the ECME process. 

Note that the VDF of the ECME-radiation electrons is shell-like, rather than being the loss-cone type, as presumed by most earlier studies(e.g., \citealt{1986Winglee}; \citealt{1988aAschwanden}; \citealt{2000Stupp}). According to our simulation, the shell-like distribution is mainly caused by the mirroring motion of energetic electrons along the converging sunspot field. We point out that the exact details of the formation of particle VDF depend on the conditions of wave-particle interaction along the loop, the shape of the initial VDF, and other factors such as the time and location of the injection.

Based on the above analysis, the spectral cutoffs of t-IVdm bursts are due to the limited ranges of altitudes ($\sim$20--100 Mm) within which most energetic electrons get mirrored. This altitude range corresponds to ECME-induced harmonic frequencies ($\sim$2 $\Omega_{ce}$) of $\sim$80 to 2000 MHz. This explains the generally well-defined spectral ranges of most t-IVdm bursts.

\section{Conclusions and Discussion} \label{sec:discussion}
Here, we investigated the radiation mechanism of flare-induced t-IVdm bursts using multi-wavelength data. We analyzed the burst on 20110924 with medium-strong levels of polarization and originating from sources near a sunspot. The emission is intermittent and the maximum $T_B$ exceeds 10$^{11}$ K, with well-defined upper and lower frequency cutoffs at $\sim$1--2 GHz and 200--300 MHz, respectively. The radio sources of the burst show left-handed polarization, and lie along the field lines pointing toward the sunspot with a negative polarity. According to the extrapolation of the coronal magnetic field, we found a good correlation between the sources and the sites of the second harmonic of the local electron gyrofrequency. These findings provide substantial evidence that the burst is the X2 radiation of ECME.

We further modeled the electron transport along the coronal loop. We found that most electrons get mirrored in converging sunspot fields within the altitude range of 20--100 Mm. This agrees with the well-defined spectral ranges of such bursts. The ECME-radiating energetic electrons exhibit a shell-like VDF, rather than the generally-presumed loss-cone distribution.
	
The modes excited by ECME depend on the ratio of plasma frequency to gyrofrequency ($\omega_{pe}/\Omega_{ce}$). Note that most of earlier studies considered the loss-cone VDF that is quite different from the shell-like VDF obtained by our modeling of electron transport along the selected loop. Earlier studies show that both O and X modes can be effectively excited for 0.8$<$$\omega_{pe}/\Omega_{ce}$$<$2.0 \citep{2000Stupp}, with O1 dominant for $\omega_{pe}/\Omega_{ce}$$<$0.5 \citep{1986Winglee} or 0.3$<$$\omega_{pe}/\Omega_{ce}$$<$1 \citep{1988aAschwanden}, X1 being dominant for $\omega_{pe}/\Omega_{ce}$$<$0.3 \citep{1988aAschwanden}, and X2 dominant for 0.5$<$$\omega_{pe}/\Omega_{ce}$$<$1.5 \citep{1986Winglee}. \citet{2021Yousefzadeh, 2022Yousefzadeh} and \citet{2021Ning} simulated the maser radiations by energetic electrons with shell-like VDFs. They found efficient excitation of X2 mode with only very weak or no excitation of O1 and X1, for $\omega_{pe}/\Omega_{ce}$$\leqslant$0.5. We also did PIC simulations for $\omega_{pe}/\Omega_{ce}$$<$1.3 to examine the ECME modes excited by energetic electrons with the VDF obtained by our particle-transport modelling and got very similar result: efficient X2 excitation while only weak or even no enhancement of O1 and X1 modes.

According to one earlier analysis \citep{1981Zlotnik}, the harmonic plasma emission could be in X mode if its mother mode, i.e., the Langmuir waves, propagates isotropically, and the polarization degree never exceeds 40\%. Yet with most updated particle-in-cell simulations (e.g., \citealt{2022Chen,2022Zhang}), the Langmuir waves driven by electron beams propagate mostly quasi-parallel to the beam. So the above result does not hold for most beam-driven situation. So we have excluded the possibility of harmonic plasma emission to account for the observed highly-polarized t-IVdm burst.

This study highlights the critical role of ECME in generating the commonly-observed wideband pulsating decimetric radio bursts. This has profound implication, not only to solar astronomy but also to stellar physics, since the ECME radiation helps in diagnosing the source field strength that remains a major challenge even in the solar case.

For the other 59 events collected here, we cannot determine the magnetic field orientation within the sources or the sources are characterized by weak polarizations. So their radiation mechanisms remain elusive. For events occurring after 2014, we cannot determine the sense of polarization because NRH stopped calibrating its polarization data since then. Further development of decimetric radio heliograph with well-calibrated data of high spatial and spectral resolution is demanded for a better understanding of such events.

\acknowledgments
The authors are grateful to the GOES, NRH, SDO/AIA, and HMI consortia for providing the data. We thank the Institute for Data Science FHNW Brugg/Windisch, Switzerland, for providing the e-Callisto data. SDO is a mission of NASA's Living With a Star Program. 
This work was supported by the National Natural Science Foundation of China (12303061, 12127901, 12333009, and 12203031), the National Key R\&D Program of China (2022YFF0503002 and 2022YFF0503000), the Shandong Natural Science Foundation of China (ZR2023QA074), and the China Postdoctoral Science Foundation (2023T160385, 2022M711931, and 2022TQ0189). A.K. was supported by the Ministry of Science and Higher Education of the Russian Federation.
We acknowledge the computational resources provided by the Beijing Super Cloud Computing Center (BSCC, URL: http://www.blsc.cn).

\bibliographystyle{aasjournal}
\bibliography{ms}

\newpage

\startlongtable
\begin{longrotatetable}
	\begin{deluxetable*}{@{\extracolsep{1pt}} p{2cm} p{2cm} p{2cm} p{1.5cm} p{1.5cm} p{2.2cm} p{2.2cm} p{2.5cm} p{2.2cm} p{2.5cm} p{0cm}}
		\tablecolumns{12}
		\centerwidetable
		\centering
		\tablecaption{Main information of the 37 flare-associated t-IVdm bursts}
		\label{tab1}
		
		\tablehead{
			No. Date & Time (UT) & Sunspot Polarity & Active Region & Flare Class & Location of Sunspot\tablenotemark{1} (arcsecond) & Location of NRH Source at 445 MHz\tablenotemark{2} (arcsecond) & Frequency Range (GHz)  & $T_{B}$ Max. (K) & Polarization Max.\tablenotemark{3} (\%)
		}
		
		\startdata
		01. 20100814 & 09:40-10:50 & Leading (Positive) Following (Negative) & 11099 & C4.4 & (780, 200) & (800, 50) & $\sim$0.1--3.0 & 4.4 $\times$ 10$^9$ (150 MHz) & \parbox[t]{3cm}{72 \\ (360 MHz)} \\
		\hline
		02. 20110307 & 14:16-14:45 & Leading (Negative) & 11166 & M1.7 & (-300, 250) & (-150, 350) & $\sim$0.1--$>$2.0 & 1.2 $\times$ 10$^{10}$ (150 MHz) & \parbox[t]{3cm}{70 \\ (360 MHz) \\ -100 \\ (445 MHz)} \\
		\hline
		03. 20110809 & 08:31-08:56 & Leading (Negative) Following (Positive) & 11263 & X6.9 & (850, 250) & (910, 80) & $\sim$0.2--0.6 & 2.2 $\times$ 10$^9$ (270 MHz) & \parbox[t]{3cm}{66 \\ (150 MHz)} \\
		\hline
		04. 20110924 & 12:50-13:50 & Leading (Negative) & 11302 & M7.1 & (-730, 120) & (-730, 100) & $\sim$0.2--$>$2.0 & 2.0 $\times$ 10$^{11}$ (327 MHz) & \parbox[t]{3cm}{-95 \\ (432 MHz)} \\
		\hline
		05. 20110925 & 09:40-10:40 & Single (Positive) & 11299 & M1.5 & (780, -500) & (950, -490) & $\sim$0.2--1.2 & 3.8 $\times$ 10$^8$ (327 MHz) & \parbox[t]{3cm}{69 \\ (445 MHz)} \\
		06. 20111005 & 11:14-11:24 & Leading (Positive) Following (Negative) & 11313 & C6.1 & (-900, -250) & (-980, -360) & $\sim$0.2--1.2 & 2.5 $\times$ 10$^{8}$ (432 MHz) & \parbox[t]{3cm}{-65 \\ (408 MHz)} \\
		\hline
		07. 20120304 & 10:35-12:15 & Leading (Positive) Following (Negative) & 11429 & M2.0 & (-820, 330) & (-70, 420) & $\sim$0.2--$>$2.0 & 1.0 $\times$ 10$^{10}$ (360 MHz) & \parbox[t]{3cm}{91 \\ (445 MHz)} \\
		\hline
		08. 20120314 & 15:17-15:26 & Single (Positive) & 11432 & M2.8 & (-300, -250) & (-60, 320) & $\sim$0.2--$>$2.0 & 1.0 $\times$ 10$^{9}$ (228 MHz) & \parbox[t]{3cm}{-32 \\ (432 MHz)} \\
		\hline
		09. 20120418 & 12:51-13:40 & Single (Positive) & 11462 & C8.9 & (450, -350) & (590, -340) & $\sim$0.1--0.6 & 2.5 $\times$ 10$^{9}$ (150 MHz) & \parbox[t]{3cm}{100 \\ (408 MHz)} \\
		\hline
		10. 20120427 & 08:22-08:27 & Leading (Negative) & 11466 & M1.0 & (500, 250) & (600, 250) & $\sim$0.3--1.2 & 2.3 $\times$ 10$^{8}$ (445 MHz) & \parbox[t]{3cm}{-10 \\ (445 MHz)} \\
		\hline
		11. 20120507-1 & 14:07-14:20 & Single (Positive) & 11471 & M1.9 & (630, -320) & (820, -220) & $\sim$0.1--1.0 & 3.1 $\times$ 10$^{8}$ (150 MHz) & \parbox[t]{3cm}{10 \\ (270 MHz)} \\
		\hline
		12. 20120507-2 & 14:36-14:41 & Single (Positive) & 11471 & M1.9 & (630, -320) & (780, -250) & $\sim$0.2--0.5 & 2.2 $\times$ 10$^{8}$ (228 MHz) & \parbox[t]{3cm}{32 \\ (228 MHz)} \\
		13. 20120613 & 13:00-13:50 & Leading (Positive) Following (Negative) & 11504 & M1.2 & \parbox[t]{3cm}{(-240, -300) \\ \\ (-390, -250)} & \parbox[t]{3cm}{(-150, -450) \\ \\ (-400, -450)} & $\sim$0.1--0.5 & 3.0 $\times$ 10$^{9}$ (150 MHz) & \parbox[t]{3cm}{90 \\ (445 MHz) \\ -90 \\ (408 MHz)} \\
		\hline
		14. 20120614 & 13:40-15:15 & Following (Negative) & 11504 & \parbox[t]{3cm}{M1.5 \\ M1.9} & (-190, -250) & (-170, -300) & $\sim$0.1--0.5 & 2.0 $\times$ 10$^{10}$ (150 MHz) & \parbox[t]{3cm}{-95 \\ (408 MHz)} \\
		\hline
		15. 20120630 & 08:26-08:44 & No sunspot & 11514 & C4.4 &  & (-400, -210) & $\sim$0.2--0.5 & 6.8 $\times$ 10$^{8}$ (270 MHz) & \parbox[t]{3cm}{-22 \\ (432 MHz)} \\
		\hline
		16. 20120708 & 10:23-10:30 & Leading (Positive) Following (Negative) & 11515 & \parbox[t]{3cm}{M1.1 \\ C6.9 \\ C5.4} & (850, -300) & (930, -340) & $\sim$0.2--0.5 & 8.9 $\times$ 10$^{7}$ (228 MHz) & \parbox[t]{3cm}{-29 \\ (228 MHz)} \\
		\hline
		17. 20120710 & 08:37-08:45 & Leading (Positive) & 11519 & C4.5 & (-300, -300) & (-220, -310) & $\sim$0.3--0.7 & 1.7 $\times$ 10$^{8}$ (445 MHz) & \parbox[t]{3cm}{16 \\ (327 MHz)} \\
		\hline
		18. 20120816 & 12:56-13:06 & Single (Negative) & 11543 & C5.2 & (570, 300) & (550, 300) & $\sim$0.1--$>$0.9 & 1.5 $\times$ 10$^{8}$ (327 MHz) & \parbox[t]{3cm}{80 \\ (408 MHz)} \\
		\hline
		19. 20130403 & 09:38-09:45 & Leading (Negative) & 11708 & C2.7 & (80, 300) & (100, 300) & $\sim$0.1--0.8 & 9.0 $\times$ 10$^{9}$ (150 MHz) & \parbox[t]{3cm}{-90 \\ (298 MHz)} \\
		20. 20130517 & 08:50-09:08 & Following (Negative) Leading (Positive) & 11748 & M3.2 & \parbox[t]{3cm}{(-550, 190) \\ \\ (-450, 250)} & \parbox[t]{3cm}{(-620, 250) \\ \\ (-470, 420)} & $\sim$0.2--1.0 & 3.0 $\times$ 10$^{9}$ (408 MHz) & \parbox[t]{3cm}{-50 \\ (408 MHz) \\ 80 \\ (445 MHz)} \\
		\hline
		21. 20130522 & 13:10-13:17 & Single (Negative) & 11745 & M5.0 & (900, 220) & (910, 280) & $\sim$0.2--0.6 & 8.1 $\times$ 10$^{7}$ (298 MHz) & \parbox[t]{3cm}{76 \\ (445 MHz)} \\
		\hline
		22. 20130605 & 08:41-09:03 & Leading (Positive) & 11762 & \parbox[t]{3cm}{B7.0 \\ M1.3} & (700, -450) & (920, -460) & $\sim$0.2--0.5 & 1.4 $\times$ 10$^{9}$ (270 MHz) & \parbox[t]{3cm}{-73 \\ (445 MHz)} \\
		\hline
		23. 20130812 & 10:40-11:36 & Multi-sunspots & 11817 & M1.5 & (-250, -450) & (-260,-370) & $\sim$0.1--2.0 & 5.9 $\times$ 10$^{8}$ (150 MHz) & \parbox[t]{3cm}{79 \\ (445 MHz)} \\
		\hline
		24. 20131025 & 14:59-15:10 & Leading (Positive) Following (Negative) & 11882 & \parbox[t]{3cm}{C7.9 \\ X2.1} & (-850, -200) & (-890, -290) & $\sim$0.2--$>$0.9 & 1.3 $\times$ 10$^{9}$ (445 MHz) & \parbox[t]{3cm}{46 \\ (327 MHz)} \\
		\hline
		25. 20131026-1 & 11:01-11:07 & Leading (Positive) Following (Negative) & 11882 & M1.8 & (-780, -200) & (-850, -100) & $\sim$0.2--0.8 & 6.4 $\times$ 10$^{8}$ (270 MHz) & \parbox[t]{3cm}{67 \\ (445 MHz)} \\
		26. 20131026-2 & 11:15-11:41 & Leading (Positive) Following (Negative) & 11882 & M1.8 & (-780, -200) & (-820, -220) & $\sim$0.2--1.4 & 4.7 $\times$ 10$^{9}$ (445 MHz) & \parbox[t]{3cm}{94 \\ (445 MHz)} \\
		\hline
		27. 20131107-1 & 12:24-12:29 & Following (Negative) & 11890 & C5.9 & (-350, -300) & (-320, -500) & $\sim$0.3--2.6 & 1.0 $\times$ 10$^{8}$ (408 MHz) & \parbox[t]{3cm}{-15 \\ (408 MHz)} \\
		\hline
		28. 20131107-2 & 14:30-15:10 & Following (Negative) & 11890 & M2.4 & (-350, -300) & (-340, -660) & $\sim$0.3--0.5 & 1.9 $\times$ 10$^{8}$ (327 MHz) & \parbox[t]{3cm}{-100 \\ (408 MHz)} \\
		\hline
		29. 20131119 & 10:20-10:25 & Multi-sunspots & 11893 & X1.0 & (870, -220) & (960, -360) & $\sim$0.4--$>$2.0 & 1.9 $\times$ 10$^{8}$ (445 MHz) & \parbox[t]{3cm}{-20 \\ (408 MHz)} \\
		\hline
		30. 20140216-1 & 09:23-09:35 & Multi-sunspots (Positive) & 11977 & M1.1 & (-50, -50) & (50, -150) & $\sim$0.2--0.6 & 2.5 $\times$ 10$^{10}$ (408 MHz) & \parbox[t]{3cm}{100 \\ (228, 408, \\ 432 MHz)} \\
		\hline
		31. 20140216-2 & 13:50-14:05 & Single (Positive) & 11977 & C3.4 & (0, -50) & (100, -100) & $\sim$0.2--$>$0.9 & 5.0 $\times$ 10$^{9}$ (228 MHz) & \parbox[t]{3cm}{95 \\ (228 MHz)} \\
		\hline
		32. 20140319 & 12:50-12:55 & Multi-sunspots & 12012 & \parbox[t]{3cm}{C1.6 \\ C3.3} & (900, -220) & (-1150, -50) & $\sim$0.2--0.6 & 3.0 $\times$ 10$^{9}$ (228 MHz) & \parbox[t]{3cm}{15 \\ (270 MHz)} \\
		33. 20140404 & 13:36-13:45 & Single (Negative) & 12027 & C8.3 & (-370, 300) & (-400, 550) & $\sim$0.1--$>$2.0 & 7.0 $\times$ 10$^{9}$ (150 MHz) & \parbox[t]{3cm}{10 \\ (270 MHz)} \\
		\hline
		34. 20140415 & 09:16-09:26 & Single (Positive) & 12035 & C8.6 & (-420, -180) & (-400, -150) & $\sim$0.2--$>$0.9 & 1.0 $\times$ 10$^{8}$ (327 MHz) & \parbox[t]{3cm}{70 \\ (445 MHz)} \\
		\hline
		35. 20140728 & 14:01-14:06 & Leading (Positive) & 12125 & C2.4 & (-700, -260) & (-750, -200) & $\sim$0.4--$>$0.8 & 1.5 $\times$ 10$^{7}$ (445 MHz) & \parbox[t]{3cm}{95 \\ (432 MHz)} \\
		\hline
		36. 20140821 & 11:57-12:05 & Following (Positive) & 12148 & C1.2 & (-570, 0) & (-620, -50) & $\sim$0.2--0.6 & 3.0 $\times$ 10$^{8}$ (298 MHz) & \parbox[t]{3cm}{95 \\ (408 MHz)} \\
		\hline
		37. 20140918 & 08:39-08:45 & Single (Negative) & 12169 & M1.2 & (-850, 120) & (-1000, 120) & $\sim$0.1--0.8 & 3.0 $\times$ 10$^{11}$ (150 MHz) & \parbox[t]{3cm}{60 \\ (327 MHz) \\ -40 \\ (432 MHz)}  \\
		\enddata
		\tablenotetext{1}{When more than one sunspot is observed, the location of the sunspot that is closest to the radio sources is presented. When the sunspots are close to each other (e.g. they are near the limb), an average location is presented.}
		\tablenotetext{2}{The coordinates represent the average location during the event.}
		\tablenotetext{3}{When the sense of polarization shows a significant inversion during the event, the values for both senses are presented.}
	\end{deluxetable*}
\end{longrotatetable}

\end{document}